\documentclass[aps,prl,twocolumn,tightenlines,superscriptaddress,showpacs,byrevtex]{revtex4}
\usepackage{graphicx}% Include figure files
\usepackage{dcolumn}% Align table columns on decimal point
\usepackage{bm}% bold math
\usepackage{color}
\graphicspath{{eps/}}
\def\et {\sigma_t}
\def\Dz {D^{0}}
\def\Db {\overline{D}{}^{0}}

\def\tp {t^{\prime}}
\def\xp {x^{\prime}}
\def\xps {x^{\prime2}}
\def\yp {y^{\prime}}
\def\etau{\overline{\Gamma}}
\def\rsdecay{\Dz \to K^-\pi^+}
\def\wsdecay{\Dz \to K^+\pi^-}
\def\simlt{\mathrel{\lower2.5pt\vbox{\lineskip=0pt\baselineskip=0pt
          \hbox{$<$}\hbox{$\sim$}}}}
\begin{document}
\title{\boldmath Improved Constraints on $\Dz$-$\Db$ Mixing in $\wsdecay$ Decays from the Belle Detector}
\affiliation{Budker Institute of Nuclear Physics, Novosibirsk} \affiliation{Chiba University, Chiba}
\affiliation{Chonnam National University, Kwangju} \affiliation{University of Cincinnati, Cincinnati,
Ohio 45221}
%%%\affiliation{University of Frankfurt, Frankfurt}
\affiliation{Gyeongsang National University, Chinju} \affiliation{University of Hawaii, Honolulu,
Hawaii 96822} \affiliation{High Energy Accelerator Research Organization (KEK), Tsukuba}
%%%\affiliation{Hiroshima Institute of Technology, Hiroshima}
\affiliation{Institute of High Energy Physics, Chinese Academy of Sciences, Beijing}
\affiliation{Institute of High Energy Physics, Vienna} \affiliation{Institute of High Energy Physics,
Protvino} \affiliation{Institute for Theoretical and Experimental Physics, Moscow} \affiliation{J.
Stefan Institute, Ljubljana} \affiliation{Kanagawa University, Yokohama} \affiliation{Korea University,
Seoul}
%%%\affiliation{Kyoto University, Kyoto}
\affiliation{Kyungpook National University, Taegu} \affiliation{Swiss Federal Institute of Technology
of Lausanne, EPFL, Lausanne} \affiliation{University of Ljubljana, Ljubljana} \affiliation{University
of Maribor, Maribor} \affiliation{University of Melbourne, Victoria} \affiliation{Nagoya University,
Nagoya} \affiliation{Nara Women's University, Nara} \affiliation{National Central University, Chung-li}
\affiliation{National United University, Miao Li} \affiliation{Department of Physics, National Taiwan
University, Taipei} \affiliation{H. Niewodniczanski Institute of Nuclear Physics, Krakow}
%%%\affiliation{Nippon Dental University, Niigata}
\affiliation{Niigata University, Niigata} \affiliation{Nova Gorica Polytechnic, Nova Gorica}
\affiliation{Osaka City University, Osaka} \affiliation{Osaka University, Osaka} \affiliation{Panjab
University, Chandigarh} \affiliation{Peking University, Beijing}
%%%\affiliation{University of Pittsburgh, Pittsburgh, Pennsylvania 15260}
\affiliation{Princeton University, Princeton, New Jersey 08544} \affiliation{RIKEN BNL Research Center,
Upton, New York 11973} \affiliation{Saga University, Saga} \affiliation{University of Science and
Technology of China, Hefei} \affiliation{Seoul National University, Seoul} \affiliation{Shinshu
University, Nagano} \affiliation{Sungkyunkwan University, Suwon} \affiliation{University of Sydney,
Sydney NSW} \affiliation{Tata Institute of Fundamental Research, Bombay} \affiliation{Toho University,
Funabashi} \affiliation{Tohoku Gakuin University, Tagajo} \affiliation{Tohoku University, Sendai}
\affiliation{Department of Physics, University of Tokyo, Tokyo} \affiliation{Tokyo Institute of
Technology, Tokyo} \affiliation{Tokyo Metropolitan University, Tokyo} \affiliation{Tokyo University of
Agriculture and Technology, Tokyo}
%%%\affiliation{Toyama National College of Maritime Technology, Toyama}
\affiliation{University of Tsukuba, Tsukuba} \affiliation{Virginia Polytechnic Institute and State
University, Blacksburg, Virginia 24061} \affiliation{Yonsei University, Seoul}
   \author{L.~M.~Zhang}\affiliation{University of Science and Technology of China, Hefei} % USTC
   \author{Z.~P.~Zhang}\affiliation{University of Science and Technology of China, Hefei} % USTC
   \author{J.~Li}\affiliation{University of Science and Technology of China, Hefei} % USTC
   \author{K.~Abe}\affiliation{High Energy Accelerator Research Organization (KEK), Tsukuba} % KEK
   \author{K.~Abe}\affiliation{Tohoku Gakuin University, Tagajo} % TohokuGakuin
% \author{N.~Abe}\affiliation{Tokyo Institute of Technology, Tokyo} % TIT
   \author{I.~Adachi}\affiliation{High Energy Accelerator Research Organization (KEK), Tsukuba} % KEK
   \author{H.~Aihara}\affiliation{Department of Physics, University of Tokyo, Tokyo} % Tokyo
   \author{D.~Anipko}\affiliation{Budker Institute of Nuclear Physics, Novosibirsk} % BINP
% \author{K.~Aoki}\affiliation{Nagoya University, Nagoya} % Nagoya
   \author{K.~Arinstein}\affiliation{Budker Institute of Nuclear Physics, Novosibirsk} % BINP
   \author{Y.~Asano}\affiliation{University of Tsukuba, Tsukuba} % Tsukuba
% \author{T.~Aso}\affiliation{Toyama National College of Maritime Technology, Toyama} % Toyama
% \author{V.~Aulchenko}\affiliation{Budker Institute of Nuclear Physics, Novosibirsk} % BINP
% ...
% Lausanne
\collaboration{The Belle Collaboration}

\begin{abstract}
We report the results of a search for $\Dz$-$\Db$ mixing in $\wsdecay$ decays
 based on 400 fb$^{-1}$ of data accumulated by the Belle detector at KEKB.
Both assuming $CP$ conservation and allowing for $CP$ violation, we fit the decay-time
 distribution for the mixing parameters $\xp$ and $\yp$, as well as for the parameter $R_D$,
the ratio of doubly-Cabibbo-suppressed decays to Cabibbo-favored decays. The 95\% confidence level
region in the ($\xps,\yp$) plane is obtained using a frequentist method. Assuming $CP$ conservation, we
find $\xps<0.72\times10^{-3}$ and $-9.9\times10^{-3}<\yp<6.8\times10^{-3}$ at the 95\% confidence
level; these are the most stringent constraints on the mixing parameters to date. The no-mixing point
($0,0$) has a confidence level of 3.9\%. Assuming no mixing, we measure $R_D=(0.377\pm0.008\pm
0.005)\%$.
\end{abstract}
 \maketitle

The phenomenon of mixing has been observed in the $K^0$-$\overline{K}{}^{0}$ and
$B^0$-$\overline{B}{}^{0}$ systems, but not yet in the $\Dz$-$\Db$ system. The parameters used to
characterize mixing are $x\equiv\Delta m/\etau$ and $y\equiv\Delta \Gamma/(2\etau)$, where $\Delta m$
and $\Delta \Gamma$ are the differences in mass and decay width between the two neutral $D$ mass
eigenstates, and $\etau$ is the average width. The mixing rate within the Standard Model is expected to
be small \cite{Shipsey}. The largest predicted values, including long-distance effects, are of order
$|x|\simlt|y|\sim(10^{-3}-10^{-2})$, and are reachable with the current experimental sensitivity.
Observation of $|x|\gg|y|$ or $CP$ violation ($CPV$) in $\Dz$-$\Db$ mixing would constitute unambiguous
evidence for new physics.

The ``wrong-sign'' (WS) process, $\wsdecay$, can proceed either through direct
doubly-Cabibbo-suppressed (DCS) decay or through mixing followed by the ``right-sign'' (RS)
Cabibbo-favored (CF) decay $\Dz\to\Db\to K^+ \pi^-$\cite{cc}. The two decays can be distinguished by
the decay-time distribution. For $|x|,|y|\ll1$, and assuming negligible $CPV$, the decay-time
distribution for $\wsdecay$ can be expressed as
\begin{equation} \frac{dN}{dt}\propto e^{-\etau t}\left[R_D+\sqrt{R_D}\yp(\etau
t)+\frac{\xps+{\yp}^2}{4}(\etau t)^2\right],\label{eq}
\end{equation}
where $R_D$ is the ratio of DCS to CF decay rates, $\xp=x\cos{\delta}+y\sin{\delta}$,
$\yp=y\cos{\delta}-x\sin{\delta}$, and $\delta$ is the strong phase difference between the DCS and CF
amplitudes. The first (last) term in brackets is due to the DCS (CF) amplitude, and the middle term is
due to interference between the two processes. The time-integrated rate ($R_{\rm{WS}}$) for $\wsdecay$
relative to that for $\rsdecay$ is $R_D+\sqrt{R_D}\yp+(\xps+{\yp}^2)/2$.

To allow for $CPV$, we apply Eq.~(\ref{eq}) to $\Dz$ and $\Db$ separately. This results in six
observables: \{$R_D^+, {\xp}^{+2}, {\yp}^+$\} for $\Dz$ and \{$R_D^-, {\xp}^{-2}, {\yp}^-$\} for $\Db$.
$CPV$ is parameterized by the asymmetries $A_D=(R_D^+ -R_D^-)/(R_D^+ +R_D^-)$ and $A_M=(R_M^+
-R_M^-)/(R_M^+ +R_M^-)$, where $R_M^{\pm}=({\xp}^{\pm2}+{\yp}^{\pm2})/2$. $A_D$ and $A_M$ characterize
$CPV$ in DCS decays and in mixing, respectively. The observables are related to $\xp$ and $\yp$ via
\begin{eqnarray}
{\xp}^{\pm}&=&\left[\frac{1\pm A_M}{1\mp A_M}\right]^{1/4}(\xp \cos{\phi}\pm \yp \sin{\phi})\label{eq1}\\
{\yp}^{\pm}&=&\left[\frac{1\pm A_M}{1\mp A_M}\right]^{1/4}(\yp \cos{\phi}\mp \xp
\sin{\phi}),\label{eq2}
\end{eqnarray}
where $\phi$ is a weak phase and characterizes $CPV$ occurring in interference between mixed and
unmixed decay amplitudes. To avoid ambiguity, we restrict $\phi$ to the range $|\phi|<\pi/2$.

This method has been exploited in previous studies \cite{E791,CLEO,Babar,JLi,FOCUS}. In our previous
measurement based on a 90 fb$^{-1}$ data sample, the value of $\yp$ was found to be slightly positive
although compatible with zero \cite{JLi}. Here we exploit the much larger data set now available to
search for $\Dz$-$\Db$ mixing with significantly higher sensitivity.

 In this Letter we present improved results of an analysis of 400 fb$^{-1}$ of data, setting more
stringent limits on mixing and $CPV$ parameters. The data were recorded by the Belle detector at the
KEKB asymmetric-energy $e^+e^-$ collider \cite{KEKB}. The Belle detector \cite{Detector} includes a
silicon vertex detector (SVD), a central drift chamber (CDC), an array of aerogel threshold Cherenkov
counters (ACC), a barrel-like arrangement of time-of-flight scintillation counters (TOF), and an
electromagnetic calorimeter. The first 157 fb$^{-1}$ of data were taken with a 2.0 cm radius beampipe
and a 3-layer SVD, while the subsequent 243 fb$^{-1}$ were collected with a 1.5 cm radius beampipe, a
4-layer SVD, and a small-cell inner drift chamber \cite{SVD2}.

We reconstruct $\Dz$ candidates from the decay chain $D^{*+} \to \pi_s^+\Dz$, $\Dz \to
K^{\pm}\pi^{\mp}$. Here, $\pi_s$ denotes the low-momentum (slow) pion, the charge of which tags the
flavor of the neutral $D$ at production. We select $\Dz$ candidates by requiring two oppositely-charged
tracks, each with at least two SVD hits in both $r$-$\phi$ and $z$ coordinate, satisfying $K$ and $\pi$
identification selection criteria. These criteria are ${\cal{L}}>0.5$ for $K$ and ${\cal{L}}<0.9$ for
$\pi$, where $\cal{L}$ is the relative likelihood for a track to be a $K$ based on the response of the
ACC and measurements from the CDC and TOF. These criteria have efficiencies of 90\% and 94\%, and
$\pi/K$ misidentification rates of 10\% and 17\%, respectively. To reject background candidates in
which the $K$ is misidentified as $\pi$ and the $\pi$ is misidentified as $K$, we recalculate
$m_{K\pi}$ with the $K$ and $\pi$ assignments swapped and reject events with $|m^{\rm
(swapped)}_{K\pi}-m_{D^0}|<28$ MeV/$c^2$ $(\sim 4.5\sigma)$. A $D^{*+}$ candidate is reconstructed by
combining a $\Dz$ candidate with a $\pi_s$ candidate; the resulting $D^{*+}$ momentum in the $e^+e^-$
center-of-mass frame ($p_{D^*}$) is required to be $>2.7$ GeV/$c$ in order to eliminate $B\overline{B}$
events and suppress the combinatorial background.

The $\Dz$ vertex is obtained by fitting its daughter tracks. The $D^*$ vertex is taken as the
intersection of the $\Dz$ trajectory with the interaction region. We constrain $\pi_s$ to originate
from the obtained $D^*$ vertex. A good $\chi^2$ for each vertex fit is required. The $\Dz$ proper decay
time $t$ is then calculated. We require the uncertainty of the decay-time $\et$ to be less than 0.7~ps
(typically, $\sigma_t\!\sim\!0.13$~ps).

The selection criteria for particle identification, the $\chi^2$ of vertex fits, and $p_{D^*}$ are
obtained by maximizing $N_{\rm{sig}}/\sqrt{N_{\rm{sig}}+N_{\rm{bkg}}}$, where $N_{\rm{sig}}$
($N_{\rm{bkg}}$) is the expected number of WS signal (background) events estimated from data in the RS
signal (WS sideband) region. We assume $R_{\rm{WS}}=0.37\%$ \cite{JLi} in the calculation. The
optimized values of the selection criteria are similar to those used previously \cite{JLi}.

We select events satisfying 1.81 GeV/$c^2<m_{K\pi}< 1.91$ GeV/$c^2$ and $0<Q<20$ MeV, where $Q\equiv
m_{K\pi\pi_s}-m_{K\pi}-m_{\pi}$ is the kinetic energy released in the decay. About 5\% of selected
events have two or more $D^{*}$ candidates associated with a single $\Dz$ candidate. If these $D^{*}$
candidates have opposite sign, the event is rejected; this reduces random $\pi_s$ background (see
below) by 30\% while reducing the signal by only 1\%. If the $D^{*}$ candidates have the same sign,
then we choose the candidate that has the best $\chi^2$ resulting from the vertex fit.

\begin{figure}[!hbtp]
\center
\includegraphics[bb=0 30 540 520, width=0.24\textwidth]{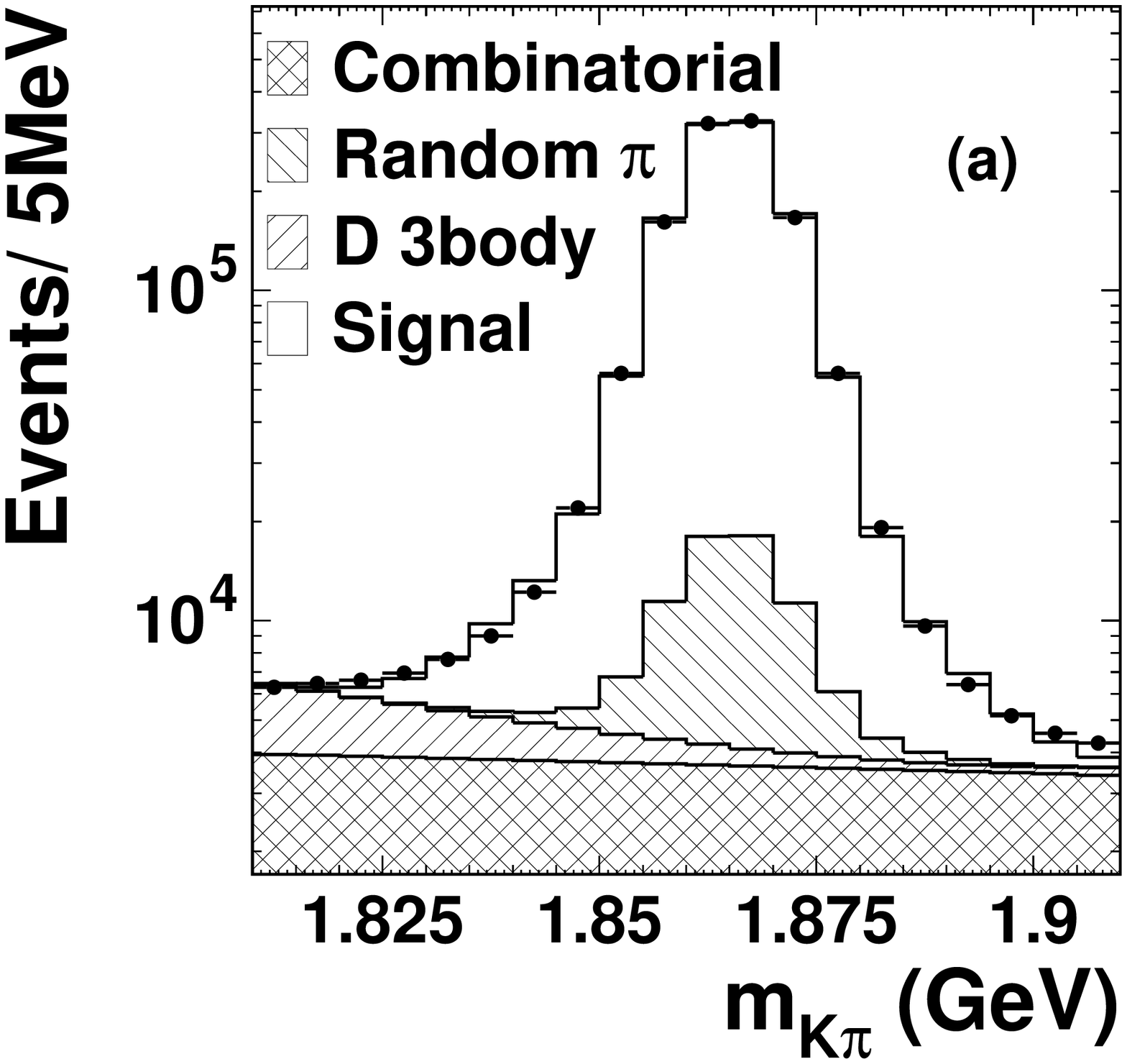}%
\includegraphics[bb=0 30 540 520,width=0.24\textwidth]{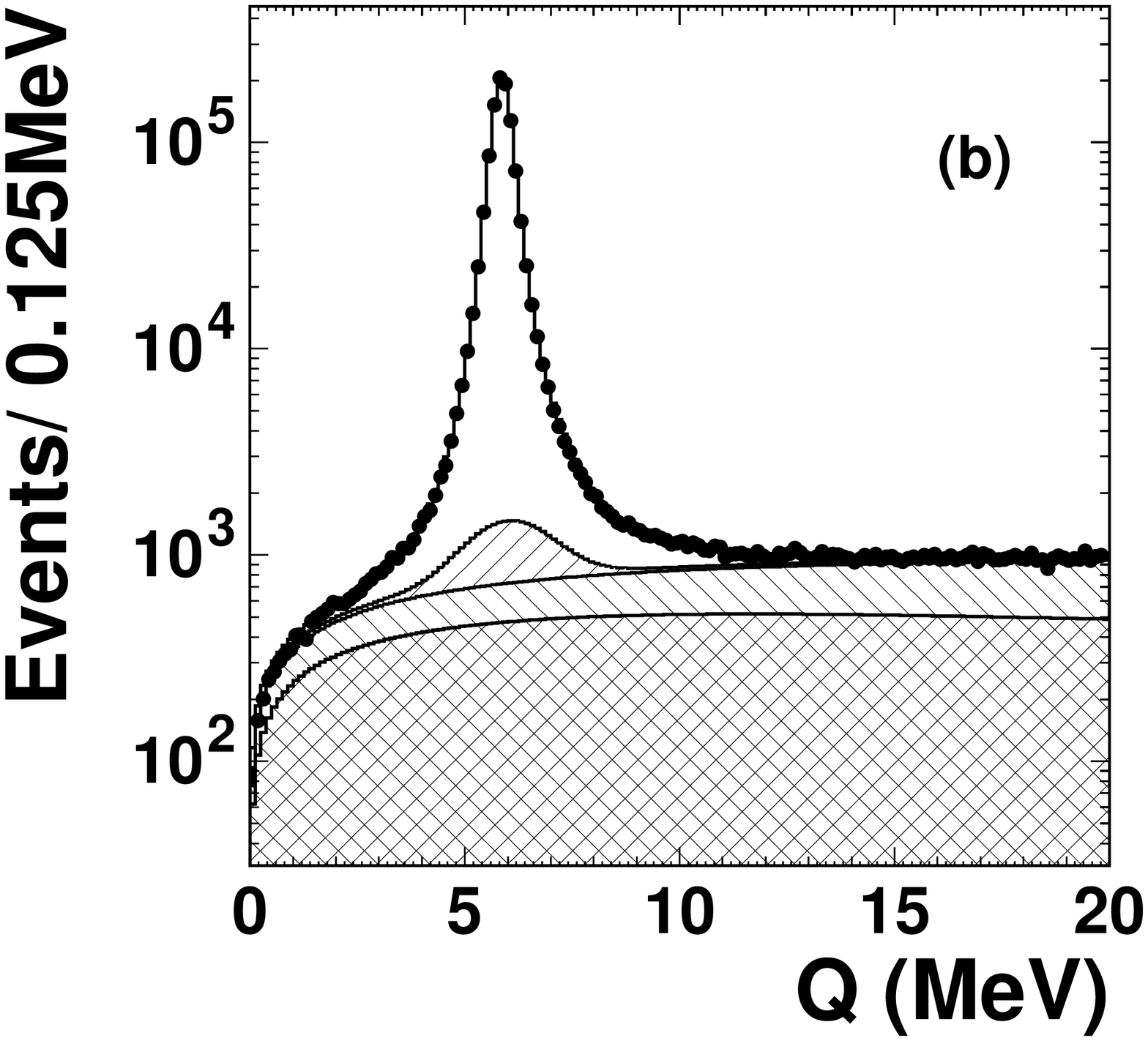}\break
\includegraphics[bb=0 30 540 520,width=0.24\textwidth]{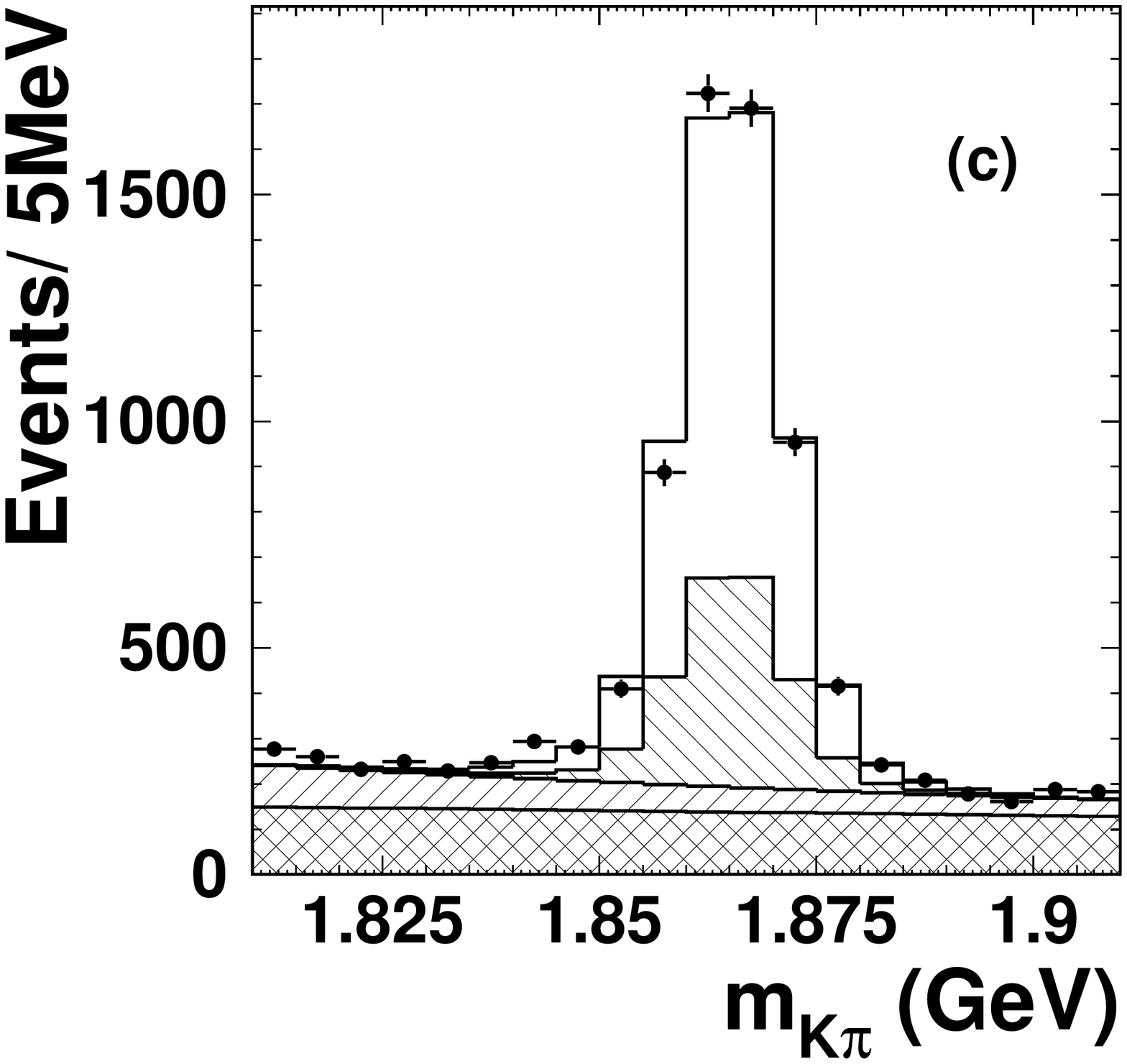}%
\includegraphics[bb=0 30 540 520,width=0.24\textwidth]{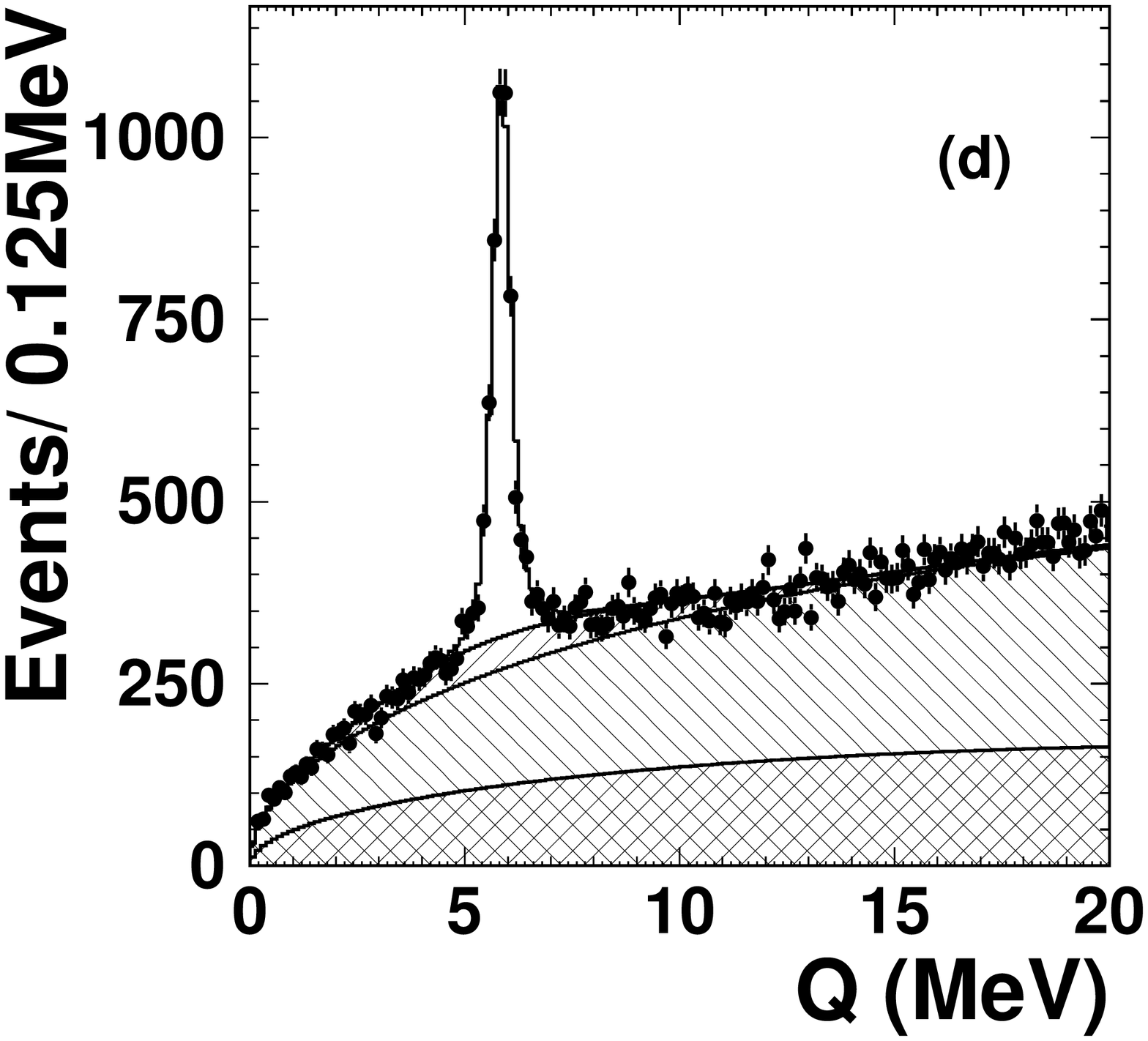}
\caption{\label{mqfig}The distribution for (a) RS $m_{K\pi}$ with $0<Q<20$ MeV; (b) RS $Q$ with 1.81
GeV/$c^2<m_{K\pi}< 1.91$ GeV/$c^2$; (c) WS $m_{K\pi}$ with $5.3$ MeV$<Q<6.5$ MeV; and (d) WS $Q$ with
1.845 GeV/$c^2$$<m_{K\pi}<1.885$ GeV/$c^2$. Superimposed on the data (points with error bars) are
projections of the $m_{K\pi}$-$Q$ fit.}
\end{figure}

We determine RS and WS event yields %the level of each background
from a two-dimensional fit to the $m_{K\pi}$-$Q$ distribution. There are four significant background
sources in the WS sample: (a) random $\pi_s$ background, in which a random $\pi^+$ is combined with a
$\Db \to K^+ \pi^-$ decay; (b) $D^{*+}\to \Dz \pi^+$ followed by $\Dz$ decaying to $\ge$ 3-body final
states; (c) $D_{(s)}^+$ decays; and (d) combinatorial. They are denoted as rnd, d3b, ds3 and cmb in
turn. These background shapes are obtained from Monte Carlo (MC) simulation and fixed in the fit. When
fitting the RS sample, the parameters for the signal shape are floated; when fitting the WS sample,
these parameters are fixed to the values obtained from the RS fit. We find $1073993\pm1108$ RS and
$4024\pm88$ WS signal events, and the ratio of WS to RS events is $(0.375\pm0.008)\%$ (statistical
error only). The ratio of WS signal to background is 1.1, about 20\% higher than that of our previous
study \cite{JLi}. The background is composed mostly of random $\pi_s$ (51\%) and combinatorial (35\%)
events. Figure~\ref{mqfig} shows the $m_{K\pi}$ and $Q$ distributions superimposed with projections of
the fit result. The WS projections
%(where the background level is higher than that for RS sample)
for $m_{K\pi}$ and $Q$ are shown for a 3$\sigma$ signal interval in $Q$ and $m_{K\pi}$, respectively.
The contribution of $D_{(s)}^+$ decays is too small to be seen.

The decay-time fitting procedure is similar to that of our previous measurement \cite{JLi} but with
several improvements as discussed below. We determine $R_D$, $\xps$ and $\yp$ by applying an unbinned
maximum likelihood fit to the distribution of WS proper decay time, considering the $4\sigma$ region
$|m_{K\pi}-m_{\Dz}|<22$ MeV/$c^2$ and $|Q-5.9$ MeV$|<1.5$ MeV. We determine background shapes by
fitting events in an $m_{K\pi}$ sideband (this contains no signal or random $\pi_s$ events). The
probability density function (PDF) for the WS signal is given by Eq.~(\ref{eq}), denoted as $P_{\rm
sig}$, convolved with a resolution function $R_{\rm sig}(t)$. The latter is represented by a sum of
three Gaussians with widths $\sigma_j=S_j \cdot \et$ ($j=1-3$) and a common mean (the error $\et$
varies event-by-event). The decay-time distributions for the backgrounds from random $\pi_s$, 3-body
and $D_{(s)}^+$ decays are exponential, $P_{k}=e^{-t/\tau_{k}}$ ($k$=rnd, d3b, ds3), while the
distribution of combinatorial background is taken to be a Dirac delta function $\delta(t)$. The
distributions are also convolved with the corresponding resolution functions $R_k$ ($k$=d3b, ds3, cmb)
which depend on $\et$. For the main background contribution of random $\pi_s$, the resolution function
and the lifetime are the same as those of the signal, since the $\pi_s$ does not affect the $\Dz$
vertex reconstruction. We define a likelihood value for each ($i$th) event as a function of $R_D$,
$\xps$ and $\yp$:
\begin{eqnarray}
\label{feqn}
&&P_i=\int_{0}^{\infty}d\tp\,\left[\left\{f_{\rm sig}^iP_{\rm sig}(\tp;R_D,\xps,\yp)\right.\right.
\nonumber\\
&+&f_{\rm rnd}^iP_{\rm rnd}(\tp)\left.\right\}
R_{\rm sig}(t_i-\tp)+f_{\rm d3b}^iP_{\rm d3b}(\tp)R_{\rm d3b}(t_i-\tp)\nonumber \\
&+&f_{\rm ds3}^iP_{\rm ds3}(\tp)R_{\rm ds3}(t_i-\tp)
\left.+f_{\rm cmb}^i\delta(\tp)R_{\rm cmb}(t_i-\tp)\right]\nonumber.\\
\end{eqnarray}
Here, the fractions $f_{k}^{i}(k=$sig, rnd, d3b, ds3 or cmb) are determined on an event-by-event basis
as functions of $m_{K\pi}$, $Q$ and $\et$.

The fitting procedure is implemented in steps as follows. First we fit the RS sideband region using a
simple background model to obtain parameters of $R_{\rm d3b}$. Then we fit the same events using a full
background model as in Eq.~(\ref{feqn}), which yields $R_{\rm cmb}$ and $\tau_{\rm d3b}$ for RS
background. We fit the RS signal region with these background parameters fixed, and obtain parameters
of $R_{\rm sig}$ (the scaling factors $S_j$, and the mean value and fractions of the individual
Gaussians) and the $\Dz$ lifetime $\tau_{\Dz}$. The latter is found to be $409.9\pm0.7$ fs, in good
agreement with the world average value \cite{PDG}. The $\chi^2$ of the fit projection on the decay-time
distribution is 64.0 for 60 bins. We use different resolution parameters for the two SVD
configurations. We then fit the WS sample. We fit the WS sideband region with $R_{\rm d3b}$ fixed from
the RS sideband fit and the $D_{(s)}^+$ contribution fixed from MC calculations; this yields $R_{\rm cmb}$ and
$\tau_{\rm d3b}$ for WS background. Finally, we fit to the WS signal region with these background
parameters, $R_{\rm sig}$ and $\tau_{\Dz}$ fixed. In the final fit, $R_D$, $\xps$ and $\yp$ are the
only free parameters and are determined by maximizing the extended log-likelihood function
$\ln{{\cal{L}}}=\sum_i\ln{P_i}+\ln{{\cal{L}}_{R}}$. The function ${\cal{L}}_{R}$ is a Gaussian that
constrains the ratio $R_{\rm{WS}}(R_D, \xps, \yp)$ to be near the value obtained from the
$m_{K\pi}$-$Q$ fit; this is needed because $P_{\rm sig}(\tp;R_D,\xps,\yp)$ is normalized to unity.

The main improvements in the decay-time fitting procedure with respect to that of our previous
measurements on a smaller data set \cite{JLi} consist of using an improved resolution function and
optimized coefficients $f_k^i$. For the latter, we include a dependence on $\et$, as this variable
substantially improves the discrimination between signal decays and combinatorial background. We
determine the $\et$ distribution for combinatorial background by fitting WS data. To check the
correctness of this method we generate MC samples with the same size as data, add the corresponding
amount of backgrounds, and repeat the fitting procedure. For a wide range of ($\xps, \yp$) values, the
fit recovers the input values well within the statistical uncertainty. If the $\et$ dependence is not
included in $f^i_k$, the fit obtains values shifted by 0.5--1 statistical standard deviation with
respect to the input values.

\begin{table}[!hbtp]
\center \caption{\label{fr}Summary of results including systematic errors. }
\begin{tabular}{lccl}\hline\hline
Fit case&Parameter&Fit result&95\% C.L. interval\\
&&($\times10^{-3}$)&($\times10^{-3}$)\\\hline
No $CPV$&$R_D$&$3.64\pm0.17$ &(3.3, 4.0)\\
&$\xps$&$0.18^{+0.21}_{-0.23}$&$<0.72$\\
&$\yp$&$0.6^{+4.0}_{-3.9}$&($-9.9$, $6.8$)\\
&$R_M$&-&($0.63\times10^{-5}$, 0.40)\\\hline
$CPV$&$A_D$&$23\pm47$&($-76$, 107)\\
&$A_M$&$670\pm1200$&($-995$, 1000)\\
%$R_D$&- &(3.3, 4.0)\\
&$\xps$&-&$<0.72$\\
&$\yp$&-&($-28$, 21)\\
&$R_M$&-&$<0.40$\\
\hline
%&$|\phi|$($^0$)&$9.4(84.5)\pm25.3$&No limits\\
No mixing&$R_D$&\multicolumn{2}{c}{$3.77\pm0.08({\rm stat.})\pm0.05({\rm syst.})$} \\\hline \hline
\end{tabular}
\end{table}

\begin{figure}[!hbtp]
\center
\includegraphics[width=0.475\textwidth]{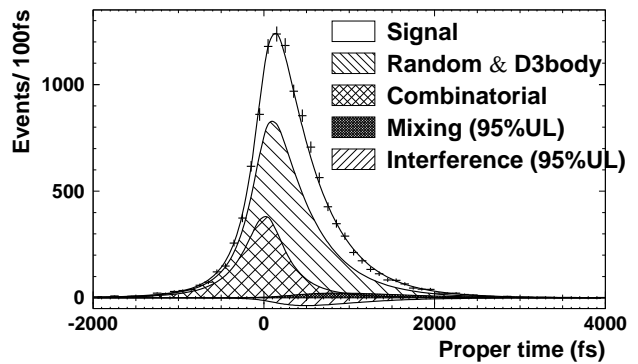}
\caption{\label{tfig}The decay-time distribution for WS events satisfying $|m_{K\pi}-m_{\Dz}|<22$
MeV/$c^2$ and $|Q-5.9|<1.5$ MeV. Superimposed on the data (points with error bars) are projections of
the decay-time fit when no $CPV$ is assumed. The mixing and interference terms are shown at the 95\%
confidence level upper limit (95\% UL) for mixing.}
\end{figure}

Table~\ref{fr} lists the results from three separate fits. For the first fit, we require $CP$ to be
conserved. The projection of this fit superimposed on the data is shown in Fig.~\ref{tfig}; the
$\chi^2$ of the projection is 54.6 for 60 bins. The central value of $\xps$ is in the
physically-allowed region $\xps>0$. The correlation between $\xps$ and $\yp$ is $-0.909$. The results
for the two SVD subsamples are consistent within 0.6 $\sigma$. For the second fit, we allow $CPV$ and
fit the WS $\Dz$ and $\Db$ samples separately. We calculate $A_D$ and $A_M$ (see Table~\ref{fr}), and
solve for $\xps$, $\yp$ and $\phi$ using Eqs.~(\ref{eq1}) and (\ref{eq2}). We obtain
$|\phi|=(9.4\pm25.3)^{\circ}$ or $(84.5\pm25.3)^{\circ}$ for the same or opposite signs of ${\xp}^+$
and ${\xp}^-$. Finally, for the last fit we assume no mixing and set $\xps=\yp=0$.

\begin{figure}[!hbtp]
\center
\includegraphics[width=0.475\textwidth]{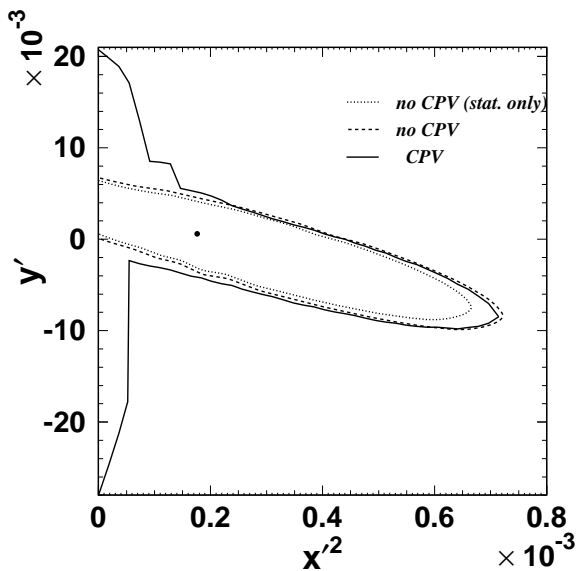}
\caption{\label{cnt} 95\% C.L. regions for ($\xps,\yp$). The point is the best fit result assuming $CP$
conservation. The dotted (dashed) line is the statistical (statistical and systematic) contour for no
$CPV$. The solid line is the statistical and systematic contour in the $CPV$-allowed case.}
\end{figure}

We apply the method described in Ref. \cite{JLi} to obtain the 95\% confidence level (C.L.) region and
take into account the systematic errors. Figure~\ref{cnt} shows the 95\% C.L. contours with and without
$CPV$ allowed. For the case of no $CPV$, the allowed area of $(\xps,\yp)$ values is smaller than that
of our previous measurement by a factor of 2.2. The $CPV$ contour has a complicated shape due to there
being two solutions for $(\xp,\yp)$ when solving Eqs.~(\ref{eq1}) and (\ref{eq2}), depending on the
signs of ${\xp}^{\pm}$.

We evaluate systematic errors by varying parameters used to select and fit the data within their
uncertainties. The sources of systematic error include event yields and imperfect modeling of
backgrounds and uncertainties in the decay-time PDF's. The former were estimated by changing the
selection criteria (kaon and pion identification, $\chi^2$ of vertex fits, and the $D^*$ momentum) and
thus the signal to background ratio over a significant range. The significance of an individual
systematic shift is found by calculating $m^2=-2[\ln{L(\vec{\alpha}_{\rm
new})}-\ln{L(\vec{\alpha}_{0})}]/2.3$, with $\vec{\alpha}_{\rm new}=({\xps}_{\rm new},{\yp}_{\rm new})$
denoting the result of the fit with the modified parameter and $\vec{\alpha}_0$ the result from the
default fit. The factor 2.3 corresponds to 68\% confidence in two dimensions. The largest shift occurs
for the $D^*$ momentum selection; it is found to be $m^2=0.083$. The parameters of functions fitted to
the $m_{K\pi}$ and $Q$ distributions were also varied by their corresponding uncertainties and the
decay-time fit was repeated. The resulting systematic error is found to be small. The influence of
$\et$ on the fractions $f_k^i$ is checked by obtaining the combinatorial background $\et$ PDF from the
fit to sideband events. Repeating the time fit with the modified $f_k^i$ yields $m^2=0.030$. The same
value is found when varying all of the fixed parameters entering the decay-time PDF's by their
uncertainty. Adding in quadrature the significances of all shifts due to possible systematic
uncertainties, we find the overall scaling factor $\sqrt{1+\sum{m_i^2}}=1.12$. We increase the 95\%
C.L. statistical contour by this factor to include systematic errors.

We show the contour with systematic errors included in Fig.~\ref{cnt} as a dashed line in the
$CP$-conserving case and as a solid line in the general case. In the case of no $CPV$, the no-mixing
point $\xps=\yp=0$ lies just outside the 95\% C.L. contour; this point corresponds to 3.9\% C.L. with
systematic uncertainty included. The two-dimensional 95\% C.L. intervals of parameters listed in
Table~\ref{fr} are obtained by projecting these contours onto the corresponding coordinate axes. In the
case of $CPV$, because the 95\% C.L. contour includes the point $\xps=\yp=0$, we cannot constrain
$\phi$ at this confidence level.

In summary, we have searched for $\Dz$-$\Db$ mixing and $CP$ violation in ``wrong-sign'' $\wsdecay$
decays using a 400 fb$^{-1}$ data sample. Assuming negligible $CP$ violation in the $\Dz$ system, we
obtain $\xps<0.72\times10^{-3}$ and $-9.9\times10^{-3}<\yp<6.8\times10^{-3}$ at 95\% C.L. These results
supercede our previous measurement and represent the most stringent limits on $\Dz$-$\Db$ mixing
parameters to date. The data exhibits a small preference for positive $\xps$ and $\yp$; the no-mixing
point $\xps = \yp = 0$ corresponds to a C.L. of 3.9\%.
\begin{acknowledgments}
We thank the KEKB group for excellent operation of the accelerator, the KEK cryogenics group for
efficient solenoid operations, and the KEK computer group and the NII for valuable computing and
Super-SINET network support.  We acknowledge support from MEXT and JSPS (Japan); ARC and DEST
(Australia); NSFC and KIP of CAS (contract No.~10575109 and IHEP-U-503, China); DST (India); the BK21
program of MOEHRD, and the CHEP SRC and BR (grant No. R01-2005-000-10089-0) programs of KOSEF (Korea);
KBN (contract No.~2P03B 01324, Poland); MIST (Russia); MHEST (Slovenia);  SNSF (Switzerland); NSC and
MOE (Taiwan); and DOE (USA).
\end{acknowledgments}


\begin{thebibliography}{999}
\bibitem{cc} Charge-conjugate modes are included throughout this Letter unless noted otherwise.
\end{thebibliography}
\end{document}